\documentclass[aps,pra,reprint,onecolumn,notitlepage,eqsecnum,floatfix,showkeys,nofootinbib]{revtex4-2}
\usepackage{amsmath}
\usepackage{amsbsy}
\usepackage{amsfonts}
\usepackage{bigints}
\usepackage{xcolor}
\usepackage{graphicx}
\usepackage{hyperref}
\usepackage{multirow}
\usepackage{linearb}
\usepackage{relsize}
\usepackage{amsthm}
\usepackage{multirow}
\usepackage{fancyvrb}
\usepackage{calrsfs}
\DeclareMathAlphabet\mathbfcal{OMS}{cmsy}{b}{n}

\newcommand{\bq}{\begin{eqnarray}}
\newcommand{\eq}{\end{eqnarray}}
\newcommand{\bqn}{\begin{eqnarray*}}
\newcommand{\eqn}{\end{eqnarray*}}
\newcommand{\bqs}{\begin{subequations}}
\newcommand{\eqs}{\end{subequations}}
\newcommand{\bw}{\begin{widetext}}
\newcommand{\ew}{\end{widetext}}
\newcommand{\xx}{{\boldsymbol x}}

\newcommand{\rr}{{\boldsymbol r}}
\newcommand{\zero}{{\boldsymbol 0}}

\newcommand{\nnabla}{{\boldsymbol\nabla}}

\newcommand{\calo}{{\cal O}}

\newcommand{\calh}{{\cal H}}

\begin{document}
%%%%%%%%%%%%%%%%%%%%%%%%%%%%%%%%%%%%%%%%%%%%%%%%%%%%%%%%%%%%%%%%%%%%%%%%%%%%%%
%%%%%%%%%%%%%%%%%%%%%%%%%%%%%%%%%%%%%%%%%%%%%%%%%%%%%%%%%%%%%%%%%%%%%%%%%%%%%%
%%%%%%%%%%%%%%%%%%%%%%%%%%%%%%%%%%%%%%%%%%%%%%%%%%%%%%%%%%%%%%%%%%%%%%%%%%%%%%
\title{Diagonal Pair Coulomb Density Matrix on a Sphere: a numerical study}

\author{Riccardo Fantoni}
\email{riccardo.fantoni@scuola.istruzione.it}
\affiliation{Universit\`a di Trieste, Dipartimento di Fisica, strada
  Costiera 11, 34151 Grignano (Trieste), Italy}

\date{\today}

\begin{abstract}
We tabulate, on the diagonal, the $n$th power of the primitive approximation 
for the pair Coulomb density matrix on a sphere, with up to $n=5$, and we 
discuss its 
behavior at particles contact as $n$ grows. Our results show that even on 
a curved surface quantum statistical physics is fundamentally different 
from its classical counterpart. If in the latter one needs to artificially 
introduce a regularization of the divergent attractive Coulomb potential 
preventing particles contact, for example with a hard core, in the former 
this is not necessary.
\end{abstract}

%\pacs{...}

\maketitle
\tableofcontents
%%%%%%%%%%%%%%%%%%%%%%%%%%%%%%%%%%%%%%%%%%%%%%%%%%%%%%%%%%%%%%%%%%%%%%%%%%%%%%
\section{Introduction}
%%%%%%%%%%%%%%%%%%%%%%%%%%%%%%%%%%%%%%%%%%%%%%%%%%%%%%%%%%%%%%%%%%%%%%%%%%%%%%
\label{sec:intro}

In a quantum many body (Monte Carlo) simulation of a Coulomb liquid it is 
often useful to have access to the low temperature pair Coulomb density matrix 
$\rho$. In the high temperature classical limit the {\sl primitive approximation} 
$\rho_{\rm primitive}$ becomes exact. Within Path Integral Monte Carlo (PIMC) of 
a many body system \cite{Ceperley1995} in order to access the quantum regime 
with less imaginary time discretization error it is important to use the 
{\sl pair-product inter-action}. The imaginary time $t$
is related to the temperature $T$ of the liquid by $t=\beta=1/k_BT$ with 
$k_B$ Boltzmann constant. 
The main ingredient to construct the pair-product inter-action is the 
inter-action of the pair density matrix or the {\sl pair-inter-action}. 
Starting from $\rho_{\rm primitive}$ at a high temperature $T_0$ 
we can halve the temperature simply by squaring it
\footnote{The squaring method can be used if an exact solution 
is known, so that successive squarings allow to reach  
temperatures $T_0/2^n$ for $n=1,2,3,\ldots$.}
\cite{Storer1968,Klemm1973} (see also section IV.G of Ref. \cite{Ceperley1995}).
This lower temperature pair density matrix already possesses different 
properties than the primitive approximation. In position representation,
on the diagonal
\bq
\left|\frac{\phi(\rr_1,\rr_2)}{k_BT_0}\right|=\left|\ln\left[
\frac{\rho_{\rm primitive}(\rr_1,\rr_2|\rr_1,\rr_2;1/k_BT_0)}
{\rho_0(\rr_1,\rr_2|\rr_1,\rr_2;1/k_BT_0)}\right]
\right|\xrightarrow{\rr_1\to\rr_2}\infty,
\eq
where $\phi(\rr_1,\rr_2)$ is the pair Coulomb potential between particle 1 at 
$\rr_1$ and particle 2 at $\rr_2$. On the other hand, for a repulsive Coulomb
potential, when Trotter formula holds \cite{Trotter1959}, we can reach the 
temperature $T_0=T_n/n$ sending $n\to\infty$ and $T_n\to\infty$, i.e.
through the {\sl continuum limit}
\footnote{For any finite $n$ one can treat the attractive Coulomb 
interaction as well by removing the divergence at contact with a 
proper cutoff.}. In this case we find
\bq
\left|P(\rr_1,\rr_2;1/k_BT_0)\right|=\lim_{n\to\infty}\left|\ln\left[
\frac{[\rho_{\rm primitive}]^n(\rr_1,\rr_2|\rr_1,\rr_2;n/k_BT_n)}
{[\rho_0]^n(\rr_1,\rr_2|\rr_1,\rr_2;n/k_BT_n)}\right]
\right|\xrightarrow{\rr_1\to\rr_2}\mbox{finite},
\eq
where we denote with $[\rho]^n$ the $n$th power of the pair density matrix:
$n=2$ for the {\sl square}, $n=3$ for the {\sl cube}, and so on.
Moreover a {\sl cusp condition} must hold for $P$ as the two particles
approach each other, like in the flat space case 
\cite{Kimball1973,Pollock1988,Ceperley1995}.

The same contact value smearing property holds at any temperature $T_0$. 
This property makes it of fundamental importance not to use the primitive 
approximation for example in a simulation of an electron proton binary 
mixture where the divergence of the attractive Coulomb potential between 
unlike species at contact would inevitably let the fast electrons stick 
onto the slow protons, and consequently spoil the simulation. 
In other words quantum statistical physics does not need hard core
regularization which is otherwise an essential but artificial feature in 
classical statistical physics necessary to avoid unlike particles collapse.

In this work we will show how this purely quantum smearing effect can 
be accomplished on a Sphere by approaching the $n\to\infty$ continuum
limit. We will not give the exact solution for the pair Coulomb density 
matrix as was done in flat space in Refs. 
\cite{Duru1982,Pollock1988,Vieillefosse1994,Vieillefosse1995} but we will
give an approximate numerical solution instead starting from the 
classical high temperature limit primitive approximation. If we 
were able to find an exact solution of the pair Coulomb density matrix 
on a sphere we would not need to worry about deriving it from the
primitive approximation. In absence of an analytic solution
here we propose to use the path integral discretized in imaginary 
time with timestep $\tau=1/k_BT_n$ to reach a temperature 
$T_0=1/nk_B\tau$ approaching the continuum limit $\tau\to 0$.
Of course this limit cannot be reached numerically but we will
be able to say something about the trend as $n$ increases.

We will only worry about the {\sl end-point approximation} 
(see section IV.F of Ref. \cite{Ceperley1995}) to the pair Coulomb
density matrix, i.e. the diagonal component of the 
pair-inter-action. Even if this is exact in the Euclidean three dimensional 
space due to the peculiar properties of the Coulomb pair potential 
\cite{Pollock1988} it becomes just an approximation on the Sphere. 

%%%%%%%%%%%%%%%%%%%%%%%%%%%%%%%%%%%%%%%%%%%%%%%%%%%%%%%%%%%%%%%%%%%%%%%%%%%%%%
\section{Approach to the continuum for the pair Coulomb density matrix}
%%%%%%%%%%%%%%%%%%%%%%%%%%%%%%%%%%%%%%%%%%%%%%%%%%%%%%%%%%%%%%%%%%%%%%%%%%%%%%
\label{sec:lower}

We show a method to lower the temperature of the hot Pair Coulomb Density 
Matrix (PCDM) on a Sphere so to extend the work in flat (Euclidean) open 
(with open boundary conditions) space of Refs. 
\cite{Pollock1988,Vieillefosse1994,Vieillefosse1995} to a 
curved surface.

We will denote with a Greek index a coordinate component and with a Roman
index a particle label. We will adopt Einstein summation convention to tacitly
assume a sum over repeated Greek indexes. We distinguish between upper
contravariant Greek indexes and lower covariant Greek indexes. Passing 
between the two is accomplished with a contraction with the metric tensor,
as usual. 

The metric tensor of the Sphere of radius $a$ is
\bq
||g_{\alpha\beta}||=a^2\left(
\begin{array}{cc}
1&0\\
0&\sin^2\theta
\end{array}
\right).
\eq
with $g(\rr)={\rm det}||g_{\alpha\beta}||=a^4\sin^2\theta$. We use 
polar coordinates $\rr=(r^1,r^2)=(\theta,\varphi)$ with $\theta\in[0,\pi]$ 
the polar angle and $\varphi\in[0,2\pi[$, so that $\theta=0$ is the north pole.

It is important to realize that on a curved surface the two body problem 
is generally not separable into a center of mass and a relative one body
problem (in Appendix \ref{app:es} we treat the case when one of the 
two particles has infinite mass). 
Therefore, in full generality, we will study the position 
representation $\rho(\rr_1,\rr_2|\rr'_1,\rr'_2;t)$ of the
density matrix $\rho=\exp(-\calh t)$, where $\calh$ is the Hamiltonian operator 
for the two bodies an $t$ is the imaginary time that in statistical physics 
is identified with the inverse temperature $\beta=1/k_BT$ with $k_B$
Boltzmann constant. These are initially at $\rr_1$ and $\rr_2$ and at an
imaginary time $t$ later they are at $\rr'_1$ and $\rr'_2$. 

Therefore, the PCDM on the Sphere will satisfy the Bloch equation
\bq \label{eq:H}
\frac{\partial\rho(\rr_1,\rr_2|\rr'_1,\rr'_2;t)}{\partial t}&=&
-\calh\rho(\rr_1,\rr_2|\rr'_1,\rr'_2;t),\\ \label{eq:bloch}
&=&\left(
\frac{\lambda_1}{\sqrt{g(\rr_1)}}{\nabla_1}_\alpha\sqrt{g(\rr_1)}\nabla_1^\alpha+
\frac{\lambda_2}{\sqrt{g(\rr_2)}}{\nabla_2}_\alpha\sqrt{g(\rr_2)}\nabla_2^\alpha-
\phi(\rr_1,\rr_2)\right)\rho(\rr_1,\rr_2|\rr'_1,\rr'_2;t),\\ \label{eq:bloch-ic}
\rho(\rr_1,\rr_2|\rr'_1,\rr'_2;0)&=&
\delta(\rr_1-\rr'_1)\delta(\rr_2-\rr'_2)/\sqrt{g(\rr_1)g(\rr_2)},
\eq
where $\calh$ is the Hamiltonian of the system of the two particles $i=1,2$, 
$\lambda_i=\hbar^2/2m_i$ with $m_i$ the 
mass of particle $i$, $g(\rr_i)=a^4\sin^2\theta_i$, 
$\nnabla_i=({\nabla_i}_1,{\nabla_i}_2)=
(\partial/\partial\theta_i,\partial/\partial\varphi_i)$ is the gradient 
respect to the coordinates $\rr_i=(r^1_i,r^2_i)=(\theta_i,\varphi_i)$,
$\delta$ is a two dimensional Dirac delta function, 
$\rho$ is the pair density matrix, and $\phi$ is the Coulomb pair potential.

The Coulomb pair potential for particles moving {\sl on} the surface can
be chosen as
\bq
\phi_{\rm on}(\rr_1,\rr_2)=Z_1Z_2e^2/d(\rr_1,\rr_2),
\eq
where $Z_ie$ is the charge of particle $i$ and for the distance between the 
two points $\rr_1$ and $\rr_2$, $d(\rr_1,\rr_2)$, on the Sphere we can either 
choose the geodesic distance 
\bq
d_g(\rr_1,\rr_2)=a\arccos[\cos\theta_1\cos\theta_2+\sin\theta_1\sin\theta_2
\cos(\varphi_1-\varphi_2)],
\eq
or, viewing the Sphere as embedded in the three dimensional space, the 
Euclidean distance
\bq
d_e(\rr_1,\rr_2)=2a\sin(d_g/2a).
\eq
For particles living {\sl in} the surface \cite{Fantoni19a} we may choose
\bq
\phi_{\rm in}(\rr_1,\rr_2)=-\frac{Z_1Z_2e^2}{2}\ln[\tan(d_g/2a)].
\eq
In fact the Green function of Poisson equation in the Sphere surface,
$\nabla_{1\alpha}\sqrt{g(\rr_1)}\nabla_1^\alpha G(\rr_1,\rr_2)=
-2\pi\delta(\theta_1-\theta_2)\delta(\varphi_1-\varphi_2)$, 
is $G=-\ln[\tan(d_g/2a)]/2$ as is shown in Appendix
\ref{app:in}
\footnote{One may also choose to neutralize each charge with a 
uniform surface charge density of opposite charge in which case
he should use $-\ln[\sin(d_g/2a)]$ instead \cite{Caillol81,Caillol2015}.
Note that while charge neutrality is a necessary condition for 
thermodynamic stability in a non compact space it is just another
possibility in a compact space like the sphere.}. We will call
$\phi_{\rm on}^g$, $\phi_{\rm on}^e$, $\phi_{\rm in}$ the three cases
just illustrated. We mention that a parallel distinction can be made
between the extrinsic and intrinsic curvature of a Sphere. We just
recall that the relationship between the intrinsic (scalar) curvature 
$R$ and the extrinsic curvature $H$ of the Sphere of radius $a$ is
$R=2H/a=2/a^2$.

We will measure energy in units of the Hartree $e^2/a_B$, with
$e$ the charge of the electron and $a_B=\hbar^2/m_ee^2$ Bohr radius, with   
$m_e$ the mass of the electron. And length in units of the Bohr radius
$a_B$. We will call $\mu_i=m_i/m_e$ the reduced mass of particle $i$.
Temperatures are in units of $\hbar^2/m_ea_B^2k_B$.

The free particles, $\phi\to 0$, pair density matrix can be determined from 
the work of Bastianelli and Corradini \cite{Bastianelli2017}. We will call 
their solution for the single free particle $i$, $\rho_{i0}(\rr_i|\rr'_i;t)$.
Therefore the pair free density matrix will be 
\bq
\rho_0(\rr_1,\rr_2|\rr'_1,\rr'_2;t)=
\rho_{10}(\rr_1|\rr'_1;t)\rho_{20}(\rr_2|\rr'_2;t).
\eq
Then our PCDM can be written as
\bq \label{eq:pia}
\rho(\rr_1,\rr_2|\rr'_1,\rr'_2;t)=
\rho_0(\rr_1,\rr_2|\rr'_1,\rr'_2;t)e^{-P(\rr_1,\rr_2|\rr'_1,\rr'_2;t)},
\eq
where $P$ is the {\sl pair-inter-action} \cite{Ceperley1995}.

Aim of this work is to tabulate $P(\rr_1,\rr_2|\rr'_1,\rr'_2;t)$ for the
three Coulomb cases illustrated above, $P_{\rm on}^g$, $P_{\rm on}^e$, 
$P_{\rm in}$. In the hot case, for small $t\ll 1$, for each of these cases, the 
{\sl primitive approximation} \cite{Ceperley1995}
\bq
P_{\rm primitive}\approx t\phi,
\eq
holds.

Therefore, introducing a timestep $\tau$, we can find the PCDM at all
subsequent timesteps through the following convolution path integral 
\bq \nonumber
\rho(\rr_1,\rr_2|\rr'_1,\rr'_2;n\tau)&=&
\int\rho(\rr_1,\rr_2|\rr_{1,1},\rr_{2,1};\tau)
\rho(\rr_{1,1},\rr_{2,1}|\rr_{1,2},\rr_{2,2};\tau)
\ldots\rho(\rr_{1,n-1},\rr_{2,n-1}|\rr'_1,\rr'_2;\tau)\times\\ \label{eq:squarer}
&&\prod_{k=1}^{n-1}\sqrt{g(\rr_{1,k})g(\rr_{2,k})}d\rr_{1,k}d\rr_{2,k},
\eq
where we denote with $\rr_{i,k}$ the position of particle $i$ at timeslice 
$k$, $d\rr=dr^1dr^2=d\theta d\varphi$, and we choose $\tau$ small enough so that
\bq
\rho(\rr_1,\rr_2|\rr'_1,\rr'_2;\tau)&\approx&
\rho_0(\rr_1,\rr_2|\rr'_1,\rr'_2;\tau)
e^{-\tau[\phi(\rr_1,\rr_2)+\phi(\rr'_1,\rr'_2)]/2}\\ \label{eq:pa}
&=&\rho_{10}(\rr_1|\rr'_1;\tau)\rho_{20}(\rr_2|\rr'_2;\tau)
e^{-\tau[\phi(\rr_1,\rr_2)+\phi(\rr'_1,\rr'_2)]/2},
\eq
is a good approximation. We will perform the $4(n-1)$ integrations in 
Eq. (\ref{eq:squarer}) through a simple Monte Carlo (MC) quadrature 
\cite{Kalos-Whitlock} where a MC point is the tuple
$(\rr_{1,1},\rr_{2,1},\ldots,\rr_{1,n-1},\rr_{2,n-1})$. 
It is important to stress that this is only
possible if $Z_1Z_2>0$ otherwise the Coulomb potential will be unbounded
from below and the path integral of Eq. (\ref{eq:squarer}) breaks down
\cite{Trotter1959}. 

According to the analysis of Ref. \cite{Bastianelli2017} 
we can further approximate
\bq \label{eq:rho0B}
g^{1/4}(\rr_i)\rho_{i0}(\rr_i|\rr'_i;\tau)g^{1/4}(\rr'_i)\approx
\frac{\mu_i}{2\pi\tau}\exp\left\{-\frac{\mu_i}{2\tau}\delta_{\alpha\beta}
(\xx_i-\xx'_i)^\alpha(\xx_i-\xx'_i)^\beta-
\tau\frac{V_R(x_i)+V_R(x'_i)}{2}\right\},
\eq
where $\delta_{\alpha\beta}$ is the Kronecker delta,
the normalization $2\pi\tau/\mu_i$ is just the one of a free particle of
mass $\mu_i$ on a plane and corresponds to the exact path integral
performed with an action 
$\int_0^\tau dt'\,\mu_i\delta_{\alpha\beta}\dot{x}_i^\alpha\dot{x}_i^\beta/2$.
Here $\xx=(x^1,x^2)$ are the Riemann normal coordinates which are related 
to the polar coordinates $\rr=(r^1,r^2)=(\theta,\varphi)$ through
\bq
x^1&=&ar^1\cos r^2,\\
x^2&=&ar^1\sin r^2,
\eq
$x=\sqrt{\delta_{\alpha\beta}x^\alpha x^\beta}$, and $V_R$ is the 
effective potential due to the scalar curvature of the Sphere, $R=2M^2$ with
$M=1/a$, namely
\bq
V_R(x)=-\frac{M^2}{6}-\frac{5(Mx)^2-3+[(Mx)^2+3]\cos(2Mx)}{48x^2\sin^2(Mx)},
\eq
which is equation (2.32) of Ref. \cite{Bastianelli2017}.
\footnote{This idea to map the fluid on the curved manifold with an equivalent
one on Euclidean space but subject to an effective external potential due to the
effect of the curvature has been used before even in the classical limit
\cite{Fantoni12e}.}
This effective
potential diverges to minus infinity for $Mx=\theta=\pi$. This divergence
will produce an exponential divergence in Eq. (\ref{eq:rho0B}) which can be 
numerically prevented through a cutoff. This procedure should not alter 
significantly the main picture since it amounts to flattening a 
neighbor of the south pole of the sphere. 

We may additionally introduce the following exact Feynman-Kac expression
for the one free body density matrix at time $t$
\bq
g^{1/4}(\rr_i)\rho_{i0}(\rr_i|\rr'_i;t)g^{1/4}(\rr'_i)&=&
\left\langle e^{-S_R}\right\rangle,~~~i=1,2\\
S_R&=&\int_0^t dt'\,V_R(x_i)
\eq
where $\langle\ldots\rangle$ denotes an average over all single 
particle Gaussian random walks from $\rr_i$ to $\rr'_i$ in a time $t$.
In Appendix \ref{app:rho0} we give the diagonal part
$\rho_{i0}(\rr_i|\rr_i;t)=\rho_{i0}(\zero|\zero;t)$ up to orders 
$t^8$. Our Eq. (\ref{eq:rho0}) coincides with the perturbative 
expansion (3.23) of Ref. \cite{Bastianelli2017} for $d=2$ 
spatial dimensions.

Even if in general we will be interested in the full 
$P(\rr_1,\rr_2|\rr'_1,\rr'_2;t)$ in practice in a many body simulation we 
may use the {\sl end point approximation} for the inter-action 
as shown in section IV.F of Ref. \cite{Ceperley1995} where one only needs 
the diagonal components $P(\rr_1,\rr_2|\rr_1,\rr_2;t)=P(d_g(\rr_1,\rr_2)/a;t)$. 
On the sphere the geodesic distance $d_g(\rr_1,\rr_2)$ between any two points 
$\rr_1$ and $\rr_2$ is bounded from above by $a\pi$. Due to the Sphere symmetry, 
without loss of generality, we may choose $\rr_1=(0,0)=\zero$ and 
$\rr_2=(\theta,0)$, so that $d_g=a\theta$.

On the diagonal we will have
\bq
\rho(\rr_1,\rr_2|\rr_1,\rr_2;t)=
\rho(d_g(\rr_1,\rr_2)/a;t)=
\rho_0(d_g(\rr_1,\rr_2)/a;t)
e^{-P(d_g(\rr_1,\rr_2)/a;t)},
\eq
we will then tabulate
\bq \label{eq:table}
P(m\delta;n\tau)=-\ln\left[
\frac{\rho(m\delta;n\tau)}{\rho_0(m\delta;n\tau)}\right],~~~n=1,2,3,\ldots,
~~~m=1,2,3\ldots,m_{\rm max}
\eq
for given small $\tau$ and $\delta$ so that $m_{\rm max}=\pi/\delta$. 
For an imaginary time $t_n=n\tau$ this will require a number of $4(n-1)$ 
quadratures for each $m$ to calculate 
$\rho(m\delta;n\tau)=\rho(d_g(\rr_1,\rr_2)/a;t_n)=
\rho(\rr_1,\rr_2|\rr_1,\rr_2;t_n)$ and
$\rho_0(m\delta;n\tau)=\rho_0(d_g(\rr_1,\rr_2)/a;t_n)=
\rho_0(\rr_1,\rr_2|\rr_1,\rr_2;t_n)$ for 
$m=1,2,3,\ldots,m_{\rm max}$, through 
the corresponding path integral convolution (\ref{eq:squarer}).
Each convolution requires 4 quadratures over the azimuthal angle 
$\varphi\in[0,2\pi[$ and the polar angle $\theta\in[0,\pi]$
of each particle. In any case, in a many body simulation the 
pair-product inter-action is needed at a
fixed small $t$, which determines the discretization error 
\cite{Ceperley1995}. And one needs to tabulate the PCDM just once, 
at the beginning of the simulations.

\begin{figure}[htbp]
\begin{center}
\includegraphics[width=10cm]{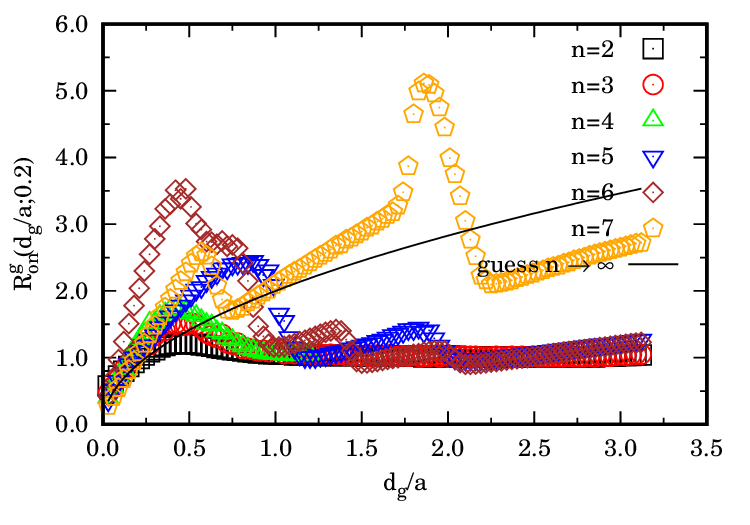}\\
\includegraphics[width=10cm]{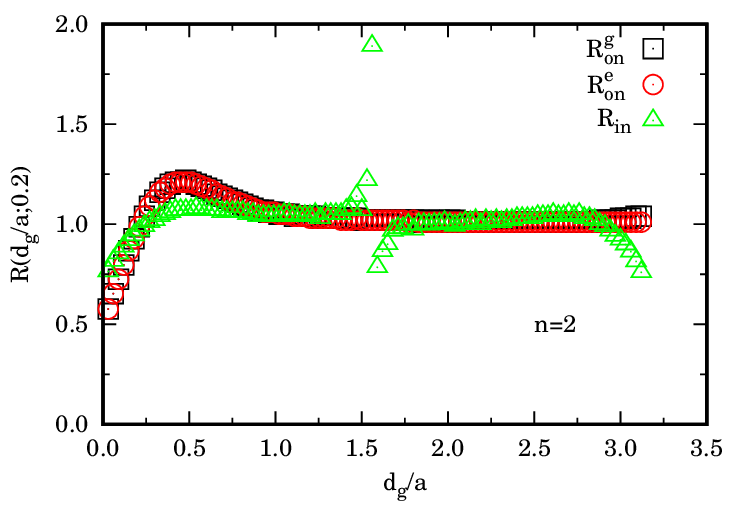}\\
\includegraphics[width=10cm]{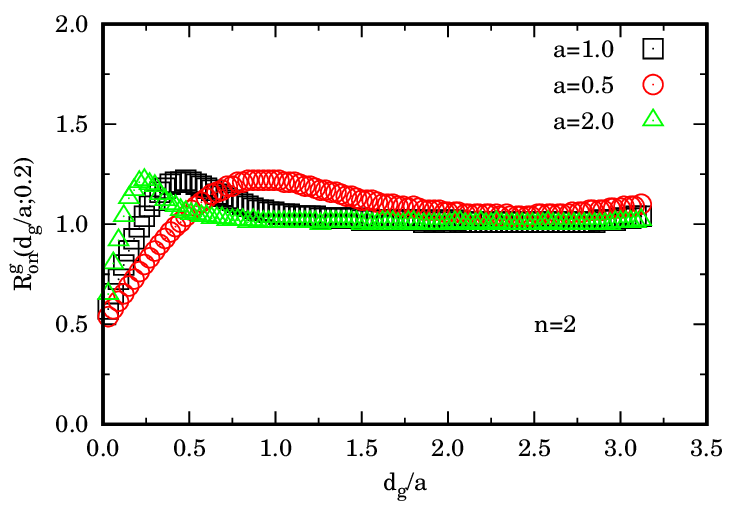}
\end{center}  
\caption{We show the plot of the ratio 
$R(d_g/a;t)=P(m\delta;t)/t\phi(m\delta)$ as a function of $d_g=m\delta$ for 
$a=1,\mu_1=\mu_2=1,Z_1=Z_2=1$, $t=n\tau=0.2,\delta=0.03$. On the top panel
$R_{\rm on}^g$ for $n=2,3,4,5,6,7$ and our guess for $n\to\infty$. For the 
cases $n=6,7$ we show only a few points. On the central panel the three 
Coulomb cases
$R_{\rm on}^g,R_{\rm on}^e,R_{\rm in}$ for $n=2$. On the bottom panel the
case $R_{\rm on}^g$ for $n=2$ and different Sphere radii. 
The angular convolutions in Eq. (\ref{eq:squarer}) were calculated with 
a Monte Carlo quadrature with no less than $10^6$ MC points and keeping 
the relative statistical error below $10\%$ 
(this required approximately $>10^{10}$ MC points in the $n>4$ cases).} 
\label{fig:ratio}
\end{figure}

In Fig. \ref{fig:ratio} we plot the ratio $R(m\delta;t)=
P(m\delta;t)/t\phi(m\delta)$ as a function of $m$ for 
$a=1,\mu_1=\mu_2=1,Z_1=Z_2=1$, $t=n\tau=0.2,\delta=0.03$ 
and $n=2,3,4,5,6,7$ in the top panel, $n=2$ in the other
two panels. In the top panel, for the cases $n=6,7$ we 
show just a few points from which we may hint the whole behavior.
We see how at large particles distance the ratio $R$
approaches 1 for $n\leq 4$ but at contact it tends to a value below 1,
approximately $\approx 1/n$ as shown in Table \ref{tab:tq}. In 
between, $R_{\rm on}^g\approx R_{\rm on}^e$, have a first peak at 
$(\bar{d}_g(n),\bar{R}_{\rm on}^g(n))$. Where 
$\bar{R}_{\rm on}^g(n)$ increases with $n$ and $\bar{d}_g(n)$
remains constant for $n\leq 4$ and grows for $n>4$ with
an exception for $n=6$. This is in 
agreement with the fact that each convolution in the path integral
(\ref{eq:squarer}) involves 4 variables (two angles for the first
particle and two for the second) and 4 convolutions
will exhaust all 4 integrations. So that for $n\leq 4$ $R$
will be smooth throughout the whole sphere.
For $n>4$, $R(d_g/a;t)$ have kinks within 
$0<d_g/a<\pi$ due to the divergence of the Coulomb potential at
contact. The behavior of the first peak as a function of $n$,
suggests that in the continuum limit $n\to\infty$ the first peak will 
reach $d_g/a=\pi$, all the kinks will be washed out, and the ratio 
$R$ will be smooth and monotonously increasing throughout the whole 
Sphere. Our results also suggest that the ratio can reach 
$R(\pi;t)>1$ in the continuum limit. Therefore the pair-inter-action
`renormalizes' the primitive approximation on the whole sphere
and in particular is free of the divergence at contact and has 
an excess interaction when the two particles are antipodal. 
For large $n\gg 1$ adding 
one more convolution will not produce any appreciable changes to the
pair density matrix of Eq. (\ref{eq:squarer}), or in other
words $\rho(\rr_1,\rr_2|\rr'_1,\rr'_2;T)$ will be to a good 
approximation a fixed point such that 
\bq
\rho(\rr_1,\rr_2|\rr'_1,\rr'_2;T)=\int \rho(\rr_1,\rr_2|\rr''_1,\rr''_2;T)
\rho(\rr''_1,\rr''_2|\rr'_1,\rr'_2;\tau)\,\sqrt{g(\rr''_1)g(\rr''_2)}
\,d\rr''_1d\rr''_2,
\eq
with $\rho(\rr_1,\rr_2|\rr'_1,\rr'_2;\tau)$ the primitive 
approximation (\ref{eq:pa}) to the high temperature density matrix. 
As we already discussed the fixed point will be a stable sink
if the pair potential is bounded from below. Therefore we expect 
that the pair-inter-action (\ref{eq:pia}) will approach the fixed
point at $t=n\tau$ and change less and less as $n$ increases.

The central panel of Fig. \ref{fig:ratio} shows the dependence
of the ratio $R$ on the choice of the Coulomb potential just
for $n=2$.

The bottom panel of the figure shows the evolution of 
the peak for $n=2$ at different values of the Sphere radius. 
It is important to note that the peak is an effect of the 
curvature of the surface and we explicitly verified that
for $n=2$ it disappears on a periodic flat square
where the ratio is a monotonous increasing function of the 
distance reaching $\approx 1$ at the edge of the square box.
This also shows the difference between a Sphere of infinite 
radius and a periodic flat square.

In Table \ref{tab:tq} we show $P(0;t)$ and $R(0;t)$ for 
$a=1,\mu_1=\mu_2=1,Z_1=Z_2=1$, $t=n\tau=0.2$ and increasing $n$.
The value at contact, $d_g=0$, is obtained through a linear 
extrapolation of the first two points at $m=1,2$ of the top panel
of Fig. \ref{fig:ratio}. We see how $R(0;t)$ decreases approximately
as $1/n$ at increasing $n$.
Therefore we expect that at constant $t=n\tau$ as $n\to\infty$,
in the $\tau\to 0$ continuum limit, the ratio $R(0;t)\to 0$.
Then we can say that $P(d_g\to 0;t)$ must be an infinite of order 
less than the infinite of $\phi(d_g\to 0)$.
Moreover the PCDM at particles contact must obey to a cusp condition 
\cite{Pollock1988} as discussed in the next section.

\begin{table}[htbp]
\caption{Table for $P_{\rm on}^g(0;t)$ and
$R_{\rm on}^g(0;t)=\lim_{\delta\to 0}
P_{\rm on}^g(\delta;t)/t\phi_{\rm on}^g(\delta)$ for 
$a=1,\mu_1=\mu_2=1,Z_1=Z_2=1$, $t=n\tau=0.2$ and increasing $n$.
The value at contact, $d_g=0$, is obtained through a linear 
extrapolation of the first two points at $m=1,2$ of the top panel
of Fig. \ref{fig:ratio}.}  
\label{tab:tq}
%\begin{ruledtabular}
\begin{tabular}{|cc|cc|}
\hline
$~~~n~~~$ & $~~~~~\tau~~~~~$ & $~~~P_{\rm on}^g(0;0.2)~~~$ & $~~~R_{\rm on}^g(0;0.2)~~~$\\ 
\hline
2 & 0.1000 &$ 5.50 $&$ 0.517 $\\
3 & 0.0666 &$ 4.31 $&$ 0.364 $\\
4 & 0.0500 &$ 3.50 $&$ 0.250 $\\
5 & 0.0400 &$ 3.10 $&$ 0.207 $\\
6 & 0.0333 &$ 3.40 $&$ 0.162 $\\
7 & 0.0286 &$ 2.33 $&$ 0.143 $\\
\hline
\end{tabular}
%\end{ruledtabular}
\end{table}

In Appendix \ref{app:code} we present the FORTRAN code used to generate
Fig. \ref{fig:ratio}.

%%%%%%%%%%%%%%%%%%%%%%%%%%%%%%%%%%%%%%%%%%%%%%%%%%%%%%%%%%%%%%%%%%%%%%%%%%%%%%
\section{Cusp condition}
%%%%%%%%%%%%%%%%%%%%%%%%%%%%%%%%%%%%%%%%%%%%%%%%%%%%%%%%%%%%%%%%%%%%%%%%%%%%%%
\label{sec:cusp}

On the sphere the cusp condition \cite{Kimball1973,Pollock1988,Ceperley1995}.
can be determined studying the asymptotic behavior as $\rr_1\to\rr_2$ of a 
vanishing {\sl residual energy} (defined in equation (4.6) of Ref. 
\cite{Ceperley1995}) for the pair-product approximation to the inter-action. 
On a Sphere as $\rr_1\to\rr_2$ we should compare
\bq
\left(
\frac{\lambda_1}{\sqrt{g(\rr_1)}}{\nabla_1}_\alpha\sqrt{g(\rr_1)}\nabla_1^\alpha+
\frac{\lambda_2}{\sqrt{g(\rr_2)}}{\nabla_2}_\alpha\sqrt{g(\rr_2)}\nabla_2^\alpha
\right)P(d_g(\rr_1,\rr_2)/a;t)=-\phi(\rr_1,\rr_2).
\eq
As $\rr_1\to\rr_2$, $d_g(\rr_1,\rr_2)\to 0$ and we are looking at a local 
property on the sphere so that we can use the result on the tangent plane 
at the point $\rr_1=\rr_2$ namely, for any $t$,
\bq
\lim_{d_g\to 0}
\frac{\partial P_{\rm on}(d_g/a;t)}{\partial d_g}&=&
\frac{Z_1Z_2}{\lambda_1+\lambda_2},\\
\lim_{d_g\to 0}
\frac{\partial P_{\rm in}(d_g/a;t)}{\partial d_g}&=&
0,
\eq
where we used the following expression for the Laplacian in an Euclidean space 
of dimension $d$ in an isotropic state
\bq
\nnabla_r^2=\frac{\partial^2}{\partial r^2}+\frac{d-1}{r}\frac{\partial}{\partial r}.
\eq

%%%%%%%%%%%%%%%%%%%%%%%%%%%%%%%%%%%%%%%%%%%%%%%%%%%%%%%%%%%%%%%%%%%%%%%%%%%%%%
\section{Conclusions}
%%%%%%%%%%%%%%%%%%%%%%%%%%%%%%%%%%%%%%%%%%%%%%%%%%%%%%%%%%%%%%%%%%%%%%%%%%%%%%
\label{sec:conclusions}

We tabulated the (end-point approximation) diagonal inter-action for the 
repulsive $Z_1Z_2>0$ pair Coulomb density matrix on the Sphere 
at a fixed temperature $T_0=T_n/n$ obtained from the classical high 
temperature primitive approximation through a discretized path integral 
calculation with imaginary timestep $\tau=1/T_n\to 0$ in the continuum
$n\to\infty$ limit. We do this from a path integral calculated through 
a Monte Carlo quadrature. This only allows the determination of the 
repulsive ($Z_1Z_2>0$) pair-inter-action because in the attractive ($Z_1Z_2<0$) 
case the primitive pair-inter-action, i.e. the bare Coulomb 
pair-potential, is unbounded from below and Trotter formula breaks down. 
We took care explicitly of the 
$n=2$ {\sl square}, the $n=3$ {\sl cube}, up to the $n=7$ case. 
Higher $n$ cases become progressively more and more computationally demanding. 
We found that the ratio between the diagonal inter-action and its primitive 
approximation tends to $\approx 1/n$ at particles contact. This suggests
that it will go to zero in the continuum limit $n\to\infty$ for a fixed
temperature. If, for the case of two particles on a periodic square
on a plane, the ratio have a monotonously increasing behavior from 0
to 1 for all $n$, on the Sphere the ratio has a monotonous behavior only 
in the continuum limit and it can reach a value bigger than 1 at the 
maximum geodesic distance between the two particles.

We also derive the cusp condition for the pair Coulomb density matrix
on the Sphere. This has the same form as in flat space since as the
two particles comes together at a point $\rr$ their diagonal pair Coulomb
density matrix reduces to the one on the tangent plane at $\rr$.

For any finite $n$ one can treat the attractive $Z_1Z_2<0$ Coulomb 
component provided some caution is taken numerically (see Appendix
\ref{app:code}). This amounts to introduce a numerical cutoff, imposing
$d_g(\rr_1,\rr_2)>\sigma$, or, in physical terms, a hard core between 
the two unlike particles. For the repulsive component we do not have 
this problem. We showed, numerically on a Sphere, the progressive 
smearing property of the divergence at contact of the pair-inter-action 
obtained taking the $n$th powers of the primitive approximation for 
increasing $n$. 
Our results shows how quantum statistical 
physics is fundamentally different from classical statistical physics: 
if in the latter one needs to artificially introduce a regularization 
of the divergent Coulomb potential preventing particles contact, for 
example with a hard core, in the former this is not necessary since
the divergences at contact in the Coulomb pair-inter-action are washed 
out.

If one is looking for the exact attractive $Z_1Z_2<0$ component of the
pair Coulomb density matrix it is not possible to derive it through a
path integral from the primitive approximation although one can gain 
numerically some insight by progressively reducing the extent $\sigma$ 
of the hard core and increasing the number $n$ of timeslices.

Even if we are still far from an exact analytic solution of the Hydrogen 
atom problem on the surface of a Sphere with the path integral method, akin
to what was done in Euclidean space in Ref. \cite{Duru1982}, nonetheless
we found some numerical results that can give some insight in the search
for a solution of the problem. We plan, in the future, to find the 
diagonal pair-inter-action near the continuum limit $n\gg 1$, up to an 
additive constant, using the path integral Monte Carlo method 
\cite{Ceperley1995} to calculate the histogram
\bq \label{eq:hist}
\rho(d/a;n\tau)\propto\langle\delta[d_g(\rr_1,\rr_2)-d]\rangle_T,
\eq
where $T=n\tau$, $\delta[\ldots]$ is a Dirac delta function, and 
$\langle\ldots\rangle$ denotes an average with respect to the 
diagonal pair density matrix (\ref{eq:squarer}) at a 
temperature $T$. Even though this calculation cannot 
determine the additive constant, this is not needed in a 
many body Monte Carlo simulation since it cancels in the determination
of any diagonal observable \cite{Ceperley1995}. Then, for example, one 
could in principle solve the problem of the electron-proton hydrogenic 
plasma on a sphere with a two step Monte Carlo strategy where in the 
first step one determines the 3 diagonal pair-inter-action 
components through (\ref{eq:hist}) and later uses them in the many body 
Monte Carlo simulation though the end-point approximation for the 
pair-product action as shown in section IV.F of Ref. \cite{Ceperley1995}
and in Ref. \cite{Fantoni26y} in flat Euclidean space.

\appendix
%%%%%%%%%%%%%%%%%%%%%%%%%%%%%%%%%%%%%%%%%%%%%%%%%%%%%%%%%%%%%%%%%%%%%%%%%%%%%%
\section{Coulomb potential {\sl in} the sphere surface}
%%%%%%%%%%%%%%%%%%%%%%%%%%%%%%%%%%%%%%%%%%%%%%%%%%%%%%%%%%%%%%%%%%%%%%%%%%%%%%
\label{app:in}

Place one particle on the north pole $\theta=0$ so that the geodesic and 
Euclidean distances between the two particles are respectively
$d_g=a\theta$ and $d_e=2a\sin(\theta/2)$. The Poisson equation in the
sphere surface is
\bq \label{eq:Poisson}
\Delta G=-2\pi\delta^{(2)},
\eq
where $\Delta=g^{-1/2}\nabla_\alpha g^{1/2}g^{\alpha\beta}\nabla_\beta$ is
the Laplace-Beltrami operator,
$\delta^{(2)}=g^{-1/2}\delta(\theta)\delta(\varphi)$ is the Dirac delta, 
and $G$ is the Coulomb potential.

Due to rotational symmetry around the $z$ axis, Poisson equation reduces
to
\bq
\partial_\theta~\sin\theta~\partial_\theta~~G(\theta)=-2\pi\delta(\theta).
\eq
For $\theta\neq 0$ a solution is
\bq
G(\theta)=-C\ln[\tan(\theta/2)],
\eq
To determine the constant $C$ we integrate Poisson equation 
(\ref{eq:Poisson}) over the sphere surface
\bq
-2\pi=\int\Delta G~g^{1/2}~dS=2\pi\left(
\left.g^{1/2}\nabla^\theta G\right|_{\theta=\pi}+
\left.g^{1/2}\nabla^\theta G\right|_{\theta=0}\right)=-4\pi C, 
\eq
where $dS=d\theta d\varphi$ and $g^{1/2}\nabla^\theta G = -C$.

Then
\bq
G(\theta)=-\frac{1}{2}\ln[\tan(\theta/2)].
\eq

%%%%%%%%%%%%%%%%%%%%%%%%%%%%%%%%%%%%%%%%%%%%%%%%%%%%%%%%%%%%%%%%%%%%%%%%%%%%%%
\section{Eigenfunctions of the Hamiltonian (\ref{eq:H}) for $\lambda_1\to 0$}
%%%%%%%%%%%%%%%%%%%%%%%%%%%%%%%%%%%%%%%%%%%%%%%%%%%%%%%%%%%%%%%%%%%%%%%%%%%%%%
\label{app:es}

If $m_1\to\infty$ we can easily find the eigenfunctions of the Hamiltonian 
(\ref{eq:H}) by fixing particle 1 at the north pole $\theta_1=0$. Then 
Shr\"odinger equation $\calh\psi_n=E_n\psi_n$ for the eigenfunctions 
$\psi_n(\theta_2,\varphi_2)$ becomes
\bq \label{eq:SE}
\left\{-\frac{\lambda_2}{a^2}\left[\frac{1}{\sin\theta}
\frac{\partial}{\partial\theta}\left(\sin\theta\frac{\partial}{\partial\theta}
\right)+\frac{1}{\sin^2\theta}\frac{\partial^2}{\partial\varphi^2}
\right]+\phi(\theta)\right\}\psi_n=E_n\psi_n,
\eq
where $\theta=\theta_2$, $\varphi=\varphi_2$, and the Coulomb potential
\bq \label{eq:C1}
\phi_{\rm on}^g(\theta)&=&Z_1Z_2e^2/a\theta,\\
\phi_{\rm on}^e(\theta)&=&Z_1Z_2e^2/a\sqrt{2(1-\cos\theta)},\\ \label{eq:C3}
\phi_{\rm in}(\theta)&=&-Z_1Z_2e^2\ln\left[\sqrt{2(1-\cos\theta)}a/\ell\right].
\eq
Since the Coulomb potential depends only on $\theta$ the azimuthal angle 
separates and we can look for eigenfunction of the following kind:
$\psi(\theta,\varphi)=\Theta(\theta)e^{\pm im\varphi}$, where $m$ is 
an integer, the 
magnetic quantum number. Substituting this into the Schr\"odinger
equation (\ref{eq:SE}) gives an ordinary differential equation in only
$\theta$. It is convenient the following change of variable $x=\cos\theta$.
Then we have to solve
\bq
\left\{-\frac{\lambda_2}{a^2}\left[
\frac{d}{d x}(1-x^2)\frac{d}{d x}
-\frac{m^2}{1-x^2}\right]+\phi(x)\right\}\Theta(x)=E\Theta(x),
\eq
where we are looking for a solution $\Theta$ which remains finite for
$x^2\leq 1$. For $\phi=0$ the solution is the usual one in terms of 
Legendre polynomials. For the Coulomb potentials of Eqs. 
(\ref{eq:C1})-(\ref{eq:C3}) one should look for a numerical solution.
The case of a finite sphere radius $a$ shows the effect of curvature
on the case of the plane $a\to\infty$. 

%%%%%%%%%%%%%%%%%%%%%%%%%%%%%%%%%%%%%%%%%%%%%%%%%%%%%%%%%%%%%%%%%%%%%%%%%%%%%%
\section{Diagonal free particle density matrix}
%%%%%%%%%%%%%%%%%%%%%%%%%%%%%%%%%%%%%%%%%%%%%%%%%%%%%%%%%%%%%%%%%%%%%%%%%%%%%%
\label{app:rho0}

Calling $\mu_i=m_i/m_e$ and $R=2/a^2$ the Sphere scalar curvature, we find 
from equation (3.23) of Ref. \cite{Bastianelli2017} 
\bq \nonumber
\ln\left[2\pi t\sqrt{g(\zero)}\rho_{i0}(\zero|\zero;t)/\mu_i\right]&=&
\frac{tR}{12}+\\ \nonumber
&&\frac{(tR)^2}{6!}\frac{1}{2}+\\ \nonumber
&&\frac{(tR)^3}{9!}16+\\ \nonumber
&&\frac{(tR)^4}{10!}\frac{59}{4}+\\ \nonumber
&&\frac{(tR)^5}{11!}\frac{58}{3}+\\ \nonumber
&&\frac{(tR)^6}{13!}\frac{550789}{1260}+\\ \nonumber
&&\frac{(tR)^7}{14!}\frac{46838}{45}+\\ \label{eq:rho0}
&&\frac{(tR)^8}{17!}\frac{596004707}{720}+
\calo(t^9),
\eq
where due to the Sphere isotropy 
$\sqrt{g(\zero)}\rho_{i0}(\zero|\zero;t)=\sqrt{g(\rr)}\rho_{i0}(\rr|\rr;t)$
for any $\rr$.

%%%%%%%%%%%%%%%%%%%%%%%%%%%%%%%%%%%%%%%%%%%%%%%%%%%%%%%%%%%%%%%%%%%%%%%%%%%%%%
\section{The code}
%%%%%%%%%%%%%%%%%%%%%%%%%%%%%%%%%%%%%%%%%%%%%%%%%%%%%%%%%%%%%%%%%%%%%%%%%%%%%%
\label{app:code}

We give here the FORTRAN code listing to generate the table of Eq. 
(\ref{eq:table}). To tabulate the unlike attractive $Z_1=-Z_2$ case
it is necessary to choose \verb1epsd1 not too small, say $0.1$ in 
order to prevent the overflow of the exponential of the 
attractive Coulomb potential when $d_g(\rr_1,\rr_2)<$\verb1epsd1.
\newpage
{\tiny\twocolumngrid
\begin{verbatim}
      program PCDMS
ccccccccccccccccccccccccccccccccccccccccccccccccccccccccccccccccccccccc
c      
c     Pair Coulomb Density Matrix for End-Point Approximation  
c     on a Sphere from the Primitive Approximation
c
c     Monte Carlo (MC) quadrature with relative error epsmc
c
c     0<ph<2*pi   0<th<pi   th=0 north pole
c     (ds/a)**2 = dph**2 + [sin(th)*dth]**2      
c      
c     a     = Sphere radius
c     tau   = time step
c     delta = geodesic distance step
c      
ccccccccccccccccccccccccccccccccccccccccccccccccccccccccccccccccccccccc
      implicit none
      integer*8 mni
      real*8    pi,a,delta,tau,epsmc,epsd
      parameter(mni=10)
      parameter(a=1.,delta=0.03,epsmc=3.d-1,epsd=1.d-6)
      real*8    mu1,mu2,z1,z2,ti,dth,dph
      real*8    th10,ph10,th20,ph20,dg0,de0,ccc
      real*8    th1,ph1,th2,ph2,dg,de
      real*8    th1p,ph1p,th2p,ph2p,dgp,dep
      real*8    rhoid,rho,pot,pcdm,fa,pidm,faid,pcdm2,pidm2
      real*8    rhor,rhoidr
      real*8    sigmac,sigmai,avpcdm,avpidm,avpcdm2,avpidm2
      real*8    ith1(mni),iph1(mni),ith2(mni),iph2(mni)
      integer*8 i,j,k,ni,id,nd,nmc,mnmc
      character potk*3,tabf*20

      pi=acos(-1.d0)

      write(*,*) 'give the timestep tau'
      read (*,*) tau
      write(*,*) 'mu1 and mu2 in electron mass units'
      read (*,*) mu1,mu2
      write(*,*) 'Z1 and Z2'
      read (*,*) z1,z2
      write(*,*) '# timestep iterations, n >= 1, n <=',mni
      read (*,*) ni
      ti =(ni+1)*tau
      write(*,*) 'imaginary time t = (n+1)*tau =',ti
      write(*,*) '# MC steps'
      read (*,*) mnmc
      write(*,*) 'give the kind of Coulomb potential (in,one,ong)'
      read (*,*) potk
      write(*,*) '....................................................'
      write(*,*) 'Sphere radius       a =',a
      write(*,*) 'time step         tau =',tau
      write(*,*) 'g distance step delta =',delta
      write(*,*) '....................................................'
      write(tabf,'(I2)') ni
      tabf=adjustl(tabf)
      tabf='table-'//trim(potk)//'-'//trim(tabf)//'.dat'
      open(unit=8,file=tabf,status='unknown')
      write(8,*) '#....................................................'
      write(8,*) '# Sphere radius       a =',a
      write(8,*) '# time step         tau =',tau
      write(8,*) '# mu1, mu2              =',mu1,mu2
      write(8,*) '# z1, z2                =',z1,z2
      write(8,*) '# ti                    =',ti
      write(8,*) '# kind of potential     =',potk
      write(8,*) '#....................................................'
      
      nd  = int(pi/delta)

c     loop over geodesic distance      
      do id=1,nd
         th10=id*delta
         ph10=0.
         th20=0.
         ph20=0.
         ccc=cos(th10)*cos(th20)+sin(th10)*sin(th20)*cos(ph10-ph20)
         if(ccc.gt.1.)  ccc=1.
         if(ccc.lt.-1.) ccc=-1.
         dg0=a*acos(ccc)
         de0=2*a*sin(dg0/2/a)
         pcdm = 0.
         pidm = 0.
         pcdm2= 0.
         pidm2= 0.
c     MC step
         nmc  = 0
 10      nmc=nmc+1
 30      do k=1,ni
            ith1(k)=rand()*pi
            ith2(k)=rand()*pi
            iph1(k)=rand()*2*pi
            iph2(k)=rand()*2*pi
         enddo
         th1=ith1(1)
         ph1=iph1(1)
         th2=ith2(1)
         ph2=iph2(1)
         ccc=cos(th1)*cos(th2)+sin(th1)*sin(th2)*cos(ph1-ph2)
         if(ccc.gt.1.)  ccc=1.
         if(ccc.lt.-1.) ccc=-1.         
         dg=a*acos(ccc)
         de=2*a*sin(dg/2/a)
         if(dg.lt.epsd)goto 30
         rhor=rho(potk,tau,a,mu1,mu2,z1,z2,
     &        th10,ph10,th20,ph20,th1,ph1,th2,ph2,dg,de,dg0,de0)
         rhoidr=rhoid(tau,a,mu1,mu2,
     &        th1,ph1,th2,ph2,th10,ph10,th20,ph20)
         if(rhor.eq.0..or.rhoidr.eq.0.)goto 30
         fa=rhor
         faid=rhoidr
         do j=2,ni
            th1=ith1(j-1)
            ph1=iph1(j-1)
            th2=ith2(j-1)
            ph2=iph2(j-1)
            th1p=ith1(j)
            ph1p=iph1(j)
            th2p=ith2(j)
            ph2p=iph2(j)
            ccc=cos(th1)*cos(th2)+sin(th1)*sin(th2)*cos(ph1-ph2)
            if(ccc.gt.1.)  ccc=1.
            if(ccc.lt.-1.) ccc=-1.
            dg=a*acos(ccc)
            de=2*a*sin(dg/2/a)
            ccc=cos(th1p)*cos(th2p)+sin(th1p)*sin(th2p)*cos(ph1p-ph2p)
            if(ccc.gt.1.)  ccc=1.
            if(ccc.lt.-1.) ccc=-1.
            dgp=a*acos(ccc)
            dep=2*a*sin(dgp/2/a)
            if(dg .lt.epsd)goto 30
            if(dgp.lt.epsd)goto 30
            rhor=rho(potk,tau,a,mu1,mu2,z1,z2,
     &           th1,ph1,th2,ph2,th1p,ph1p,th2p,ph2p,dg,de,dgp,dep)
            rhoidr=rhoid(tau,a,mu1,mu2,
     &           th1,ph1,th2,ph2,th1p,ph1p,th2p,ph2p)
            if(rhor.eq.0..or.rhoidr.eq.0.)goto 30
            fa=fa*rhor
            faid=faid*rhoidr
         enddo
         th1=ith1(ni)
         ph1=iph1(ni)
         th2=ith2(ni)
         ph2=iph2(ni)
         ccc=cos(th1)*cos(th2)+sin(th1)*sin(th2)*cos(ph1-ph2)
         if(ccc.gt.1.)  ccc=1.
         if(ccc.lt.-1.) ccc=-1.
         dg=a*acos(ccc)
         de=2*a*sin(dg/2/a)
         if(dg.lt.epsd)goto 30
         rhor=rho(potk,tau,a,mu1,mu2,z1,z2,
     &        th1,ph1,th2,ph2,th10,ph10,th20,ph20,dg,de,dg0,de0)
         rhoidr=rhoid(tau,a,mu1,mu2,
     &        th1,ph1,th2,ph2,th10,ph10,th20,ph20)
         if(rhor.eq.0..or.rhoidr.eq.0.)goto 30
         fa=fa*rhor
         faid=faid*rhoidr
         pcdm   = pcdm+fa
         pidm   = pidm+faid
         pcdm2  = pcdm2+fa**2.
         pidm2  = pidm2+faid**2.
         avpcdm = pcdm/nmc
         avpidm = pidm/nmc
         avpcdm2= pcdm2/nmc
         avpidm2= pidm2/nmc
         sigmac = sqrt((avpcdm2-avpcdm**2.)/nmc)/avpcdm
         sigmai = sqrt((avpidm2-avpidm**2.)/nmc)/avpidm
         if(mod(nmc,mnmc).eq.0)then
            print *,'...................Rel. Err.....................'
            print *,nmc,sigmac,sigmai
            print *,'................................................'
         endif
c         if(nmc.gt.mnmc)goto 20
         if(nmc.gt.mnmc.and.sigmac.lt.epsmc.and.sigmai.lt.epsmc)goto 20
         goto 10
c     tabulate P=-log(rho/rho0)
 20      pcdm = -log(avpcdm/avpidm)
         write(8,100) dg0,pcdm,ti*pot(potk,a,z1,z2,dg0,de0),
     &        pcdm/(ti*pot(potk,a,z1,z2,dg0,de0))    
         call flush(8)
         write(*,*)   dg0,pcdm,ti*pot(potk,a,z1,z2,dg0,de0),
     &        pcdm/(ti*pot(potk,a,z1,z2,dg0,de0))    
      enddo
      
 100  format(4(d10.3,3x))
      close(unit=8)
      
      stop
      end
      
      
      function fact(n)
      implicit none
c     factorial function        
      real*8    fact,p
      integer   i,n

      p=1.
      do i=1,n
         p=p*dble(i)
      enddo
      fact=p

      return
      end


      function vr(a,mu,th)
      implicit none
c     Eq. (2.32) Bastianelli and Corradini EPJC 77, 731 (2017)
c     choose the Riemann normal coordinatees with origin on the
c     north pole th=0
      real*8    vr,a,mu,th
      real*8    m,mx,eps
      eps=1.d-6
      
      m =1/a
      mx=th
      if(mx.eq.0.) mx=eps
      vr=-m**2./6-(5*mx**2.-3+(mx**2.+3)*cos(2*mx))/48/(mx*sin(mx)/m)**2.
c     cutoff at the south pole
      if(vr.lt.-10.) vr=-10.
      
      return
      end      

      
      function pot(potk,a,z1,z2,dg,de)
      implicit none
c     the Coulomb potential
      real*8    pot,z1,z2,a
      real*8    de,dg
      character potk*3

      if(potk.eq.'in') then
         pot=-z1*z2*log(tan(dg/2/a))/2
      elseif(potk.eq.'one') then
         pot=z1*z2/de
      elseif(potk.eq.'ong') then
         pot=z1*z2/dg
      else
         write(*,*) 'no pair-potential available !'
         stop
      endif
      
      return
      end      


      function rho(potk,tau,a,mu1,mu2,z1,z2,
     &     th1,ph1,th2,ph2,th1p,ph1p,th2p,ph2p,dg,de,dgp,dep)
      implicit none
c     primitive approximation for rho
      real*8    rho,tau,a,mu1,mu2,z1,z2,dg,de,dgp,dep
      real*8    th1,ph1,th2,ph2,th1p,ph1p,th2p,ph2p
      real*8    pot,vr,ddx1,ddx2
      character potk*3

      ddx1=(a*th1*cos(ph1)-a*th1p*cos(ph1p))**2.+
     &     (a*th1*sin(ph1)-a*th1p*sin(ph1p))**2.
      ddx2=(a*th2*cos(ph2)-a*th2p*cos(ph2p))**2.+
     &     (a*th2*sin(ph2)-a*th2p*sin(ph2p))**2.

      rho =    exp(-mu1*ddx1/2/tau-tau*
     &     (vr(a,mu1,th1)+vr(a,mu1,th1p))/2)
      rho =rho*exp(-mu2*ddx2/2/tau-tau*
     &     (vr(a,mu2,th2)+vr(a,mu2,th2p))/2)
      rho =rho*exp(-tau*(pot(potk,a,z1,z2,dg,de)+
     &     pot(potk,a,z1,z2,dgp,dep))/2)

      return
      end         


      function rhoid(tau,a,mu1,mu2,
     &     th1,ph1,th2,ph2,th1p,ph1p,th2p,ph2p)
      implicit none
c     primitive approximation for ideal gas rho
      real*8    rhoid,tau,a,mu1,mu2
      real*8    th1,ph1,th2,ph2,th1p,ph1p,th2p,ph2p
      real*8    vr,ddx1,ddx2

      ddx1=(a*th1*cos(ph1)-a*th1p*cos(ph1p))**2.+
     &     (a*th1*sin(ph1)-a*th1p*sin(ph1p))**2.
      ddx2=(a*th2*cos(ph2)-a*th2p*cos(ph2p))**2.+
     &     (a*th2*sin(ph2)-a*th2p*sin(ph2p))**2.
      
      rhoid =      exp(-mu1*ddx1/2/tau-tau*
     &     (vr(a,mu1,th1)+vr(a,mu1,th1p))/2)
      rhoid =rhoid*exp(-mu2*ddx2/2/tau-tau*
     &     (vr(a,mu2,th2)+vr(a,mu2,th2p))/2)

      return
      end        


      function rho0(a,mu,t)
      implicit none
c     EQ. (3.23) Bastianelli and Corradini EPJC 77, 731 (2017)
      real*8    rho0,a,mu,t,th
      real*8    pi,fact,rit
      real*8    c1,c2,c3,c4,c5,c6,c7,c8
      parameter(c1=1./2,c2=16.,c3=59./4,c4=58./3,c5=550789./1260,
     &     c6=46838./45,c7=596004707./720)
      pi=acos(-1.d0)

      rit=t*2./a**2.
      rho0=rit/12+c1*rit**2./fact(6)+c2*rit**3./fact(9)+
     &     c3*rit**4./fact(10)+c4*rit**5./fact(11)+c5*rit**6./fact(13)+
     &     c6*rit**7./fact(14)+c7*rit**8./fact(17)
      rho0=mu*exp(rho0)/(2.*pi*t)

      return
      end      
\end{verbatim}
}
\onecolumngrid 
%%%%%%%%%%%%%%%%%%%%%%%%%%%%%%%%%%%%%%%%%%%%%%%%%%%%%%%%%%%%%%%%%%%%%%%%%%%%%%
%%%%%%%%%%%%%%%%%%%%%%%%%%%%%%%%%%%%%%%%%%%%%%%%%%%%%%%%%%%%%%%%%%%%%%%%%%%%%%
%%%%%%%%%%%%%%%%%%%%%%%%%%%%%%%%%%%%%%%%%%%%%%%%%%%%%%%%%%%%%%%%%%%%%%%%%%%%%%

\section*{Author declarations}

\subsection*{Conflicts of interest}
None declared.

\subsection*{Data availability}
The data that support the findings of this study are available from the 
corresponding author upon reasonable request.

\subsection*{Funding}
None declared.

%%%%%%%%%%%%%%%%%%%%%%%%%%%%%%%%%%%%%%%%%%%%%%%%%%%%%%%%%%%%%%%%%%%%%%%%%%%%%%
\bibliography{pcdms}
%\bibliographystyle{prsty}

%%%%%%%%%%%%%%%%%%%%%%%%%%%%%%%%%%%%%%%%%%%%%%%%%%%%%%%%%%%%%%%%%%%%%%%%%%%%%%
%%%%%%%%%%%%%%%%%%%%%%%%%%%%%%%%%%%%%%%%%%%%%%%%%%%%%%%%%%%%%%%%%%%%%%%%%%%%%%
%%%%%%%%%%%%%%%%%%%%%%%%%%%%%%%%%%%%%%%%%%%%%%%%%%%%%%%%%%%%%%%%%%%%%%%%%%%%%%
\end{document}